\documentclass{iau}

\usepackage{amsmath}
\usepackage{graphicx}
\usepackage{multirow}
\usepackage{url}
\usepackage{xurl}
\usepackage{microtype}
\jnlDoiYr{2025}
\makeatletter
\def\doitext#1{}
\makeatother

\begin{document}

\lefttitle{Barentine}
\righttitle{}

\jnlPage{1}{7}
\jnlDoiYr{2025}

\aopheadtitle{Oxford XIII / IAU Symposium 399}
\editors{D.~Hamacher, J.~Holbrook, eds.}

\title{`Unacceptable to our people': Diverse cultural beliefs, Indigenous rights, and the future of human activities on the Moon}

\author{John C.~Barentine}
\affiliation{Dark Sky Consulting, LLC}
\affiliation{PMB 237, 9420 E. Golf Links Rd., Ste 108, Tucson, AZ 85730-1317 USA}

\begin{abstract}
The human presence in outer space is undergoing a transition from one in which nation states are the dominant actors to an emerging status quo in which states merely supervise the activities of private entities. Such largely commercial ventures include extracting natural materials from the Moon, a celestial body of great cultural and spiritual reverence for some Indigenous societies. However, the existing international legal framework governing activities in space focuses on its ``exploration and use'', centered in a Western worldview that attaches to a past history of colonialism and exploitation of people and resources. While that framework, articulated in the Outer Space Treaty (OST), claims to guarantee that outer space will remain ``the province of all [hu]mankind,'' only entities with significant political power have to date decided the limits of the acceptable uses of space. This paper examines the historical record for clues about how states, private actors and Indigenous societies might interact in the future on matters of outer space governance to achieve more just ends. It analyzes a key case study: the dispute between the U.S. National Aeronautics and Space Administration (NASA) and the Diné people of the American Southwest over the launches of human cremated remains to the Moon in 1998 and 2024, acts the Diné president called ``deeply disturbing and unacceptable to our people and many other tribal nations.'' In a future in which space becomes increasingly commercialized and entities like NASA transform into mere contract-administering agencies, it is unclear how an impending, exploitative human presence on the Moon can simultaneously honor Indigenous rights and perspectives on lunar issues. The presentation concludes that best practices for future engagement with the Moon must transcend the mere ``due regard'' and ``international consultations'' required by the OST in favor of arrangements where participants co-create a human future in outer space.
\end{abstract}


\maketitle

\section{Introduction}
We live in an era of rapidly changing human activities in outer space. The arrival of the so-called ``New Space'' era marked the beginning of a substantial shift in the number and diversity of space actors as commercial entities began to overtake nation-states as the main presence in space~\citep[1--2]{Paikowsky2017}. This raises new concerns about the status of existing space law in regard to the rights of people and groups with comparatively little political capital useful in changing policies in their favor. In particular, this paper examines the place in this changing landscape occupied by Indigenous people and their governments. It considers the situation broadly through an historical examination of two episodes that brought the U.S. government into conflict with the Diné people, also known as the Navajo, who occupy traditional lands in the U.S. states of Arizona, Utah and New Mexico, over the placement of human cremated remains on the Moon. It reveals deficiencies in the international space policy status quo that effectively ignore the opinions of Indigenous people in contravention of the Outer Space Treaty\footnote{18 U.S.T.~2410, 610 U.N.T.S.~205, 61 I.L.M.~386 (1967).} (hereinafter, `OST') requirement that space ``shall be the province of all mankind''. It further underscores a tension inherent in the Treaty between Western views of space as an object of primarily human utility, and human activities in space as transactions, versus Indigenous beliefs regarding nature as animate and possessed of certain rights to be treated with reciprocity and respect~\citep[5]{Neilson2024a}. 

The history of spaceflight began near the end of the Second World War with the launch of the first human-made objects to reach space. For more than 20 years that followed, during the period in which the United States and Soviet Union emerged as the world’s superpowers, there was no comprehensive agreement between the countries, or among other nations, governing activities in space. Both superpowers threatened use of space for purposes of warfare, and they tested nuclear weapons in space between 1961 and the entry into force of the Partial Test Ban Treaty\footnote{``Treaty Banning Nuclear Weapon Tests in the Atmosphere, in Outer Space and Under Water'', 14 U.S.T.~1313, 480 U.N.T.S.~43, A.T.S~26, 2 I.L.M.~889 (1963).}. After the U.S. announcement in 1960 that it intended to land astronauts on the Moon, concerns arose that a territorial claim would follow. These and other considerations led to the process by which the modern international space policy framework was constructed~\citep[350--372]{Buono2020}.

Yet it has been argued recently that for all of the rhetoric in the OST concerning space as ``the province of all mankind'' (Article I), it was only in part by excluding the developing world, and many Indigenous peoples, that international agreement leading to the Treaty was possible~\citep[93--94]{vanEijk2025}. Similarly, although the Moon Agreement\footnote{``Agreement Governing the Activities of States on the Moon and Other Celestial Bodies '', 1363 U.N.T.S.~22, 18 I.L.M.~1434 (1979).} refers to ``the moon and its natural resources'' as ``the common heritage of mankind'', this instrument that flows from the OST was similarly negotiated without the direct participation of Indigenous nations~\citep[25]{Christol1999}. It is against this backdrop that we consider first a series of historic controversies involving the Moon, leading up to a pair of events that exemplifies the ongoing tension between the United States and the Navajo Nation where outer space activities are concerned.

\section{Previous controversies involving the Moon}
During the height of the classic, Apollo-era Space Race, a number of controversial events took place involving the actions of NASA astronauts as employees of the U.S. federal government. On Christmas Eve 1968, astronauts aboard \emph{Apollo 8} — the first humans to reach the orbital space of the Moon — famously read from the biblical book of Genesis during a live television broadcast. Madalyn Murray O'Hair, founder of an organization called American Atheists, sued the U.S. government in federal district court alleging that the reading violated the Establishment Clause of the First Amendment to the U.S. Constitution~\citep[623]{Chaikin1994}. The case, \emph{O'Hair v. Payne},\footnote{U.S.~District Court for the Western District of Texas, 312 F.~Supp.~434 (W.D.~Tex.~1969); 397 U.S. 531 (1970).} was eventually presented to the U.S. Supreme Court, which declined to hear it. 

The following year, after landing on the Moon during \emph{Apollo 11}, astronaut Edwin E.~Aldin, Jr., celebrated communion. Given that \emph{O'Hair} was still pending at the time, NASA was wary of further religious displays in space. Although contemporaneous media stories noted that Aldrin planned to bring communion bread aboard the spacecraft, NASA advised Aldrin to keep his communion ceremony private. News of the event was not revealed publicly until after the end of the mission~\citep{Cresswell2012}.

Controversy returned during the \emph{Apollo 15} mission in 1971 due to \emph{Fallen Astronaut}, an 8.9-cm aluminum sculpture created by Belgian artist Paul Van Hoeydonck. Placed on the Moon during the mission as a tribute to astronauts and cosmonauts who died in the course of advancing space exploration, it was disclosed to the public only after the \emph{Apollo 15} crew returned to Earth. Van Hoeydonck attempted to sell up to 950 copies of the sculpture for USD 750 each, running afoul of a strict NASA policy prohibiting commercial exploitation of the publicly funded space program. Van Hoeydonck canceled his plans to sell the replicas after receiving complaints from NASA~\citep{PowellShapiro2013}. On the same mission, astronauts carried about 400 unauthorized postal covers to the Moon to be sold for profit after they returned home. When the scheme was uncovered, NASA reprimanded the astronauts for poor judgment; none flew into space again. ``For better or for worse,'' Apollo historian Andrew Chaikin wrote of the episode, ``the myth of the Perfect Astronaut had crumbled''~\citep[498]{Chaikin1994}.

Taking place near the height of the Cold War, deepening U.S. involvement in the Vietnam War, and domestic racial unrest, the Moon landings were subject to various forms of social criticism. Activists decried the financial cost of the Apollo program given myriad problems on Earth going unsolved. It was also around the dawn of the modern environmental movement, and some were already thinking ahead to preservation of extraterrestrial landscapes. Critics of \emph{Apollo} also pointed out the very presence of the astronauts on the Moon led to its literal defilement: the extensive list of refuse left behind during the program includes 96 bags of human waste jettisoned on the lunar surface~\citep{NASA2012}.

\section{1998 and 2024 incidents}
Two historical episodes in 1998 and 2024 offer insight into the ways in which the government of a spacefaring nation, the United States, dealt with objections raised by Indigenous nations with respect to its activities on the Moon. These episodes illustrate the complex interplay between the rights of Indigenous peoples, religious freedom, the OST, and the actions of commercial entities. They also point out shortcomings in the current space policy regime that make it ripe for substantial reconsideration in the new Space Age.

On 7 January 1998, NASA launched the uncrewed \emph{Lunar Prospector} spacecraft toward lunar orbit. Its mission was to map the Moon’s surface composition including potential deposits of hydrogen; measure lunar magnetic and gravity fields; and study geological outgassing events. Its main science result was the detection of hydrogen, implying the existence of ice deposits. After 570 days in orbit, the mission ended on 31 July 1999 when, consistent with NASA ``Planetary Protection'' guidelines for the disposal of spacecraft at the end of their missions, \emph{Lunar Prospector} was deliberately crashed into a crater near the lunar south pole.

A small polycarbonate capsule aboard the spacecraft contained one ounce of the cremated remains of the American planetary scientist Dr. Eugene ``Gene'' Shoemaker, who was killed in an automobile accident in July 1997. Sending the ashes to the Moon was the idea of Shoemaker’s former student, Carolyn Porco. ``It was legend among planetary scientists that Gene’s life-long dream was to go to the moon and study its geology firsthand,'' Porco said in 1999. ``At his journey’s end — thirty years to the month after humans first set foot on the moon — Eugene M. Shoemaker will become the first inhabitant of Earth to be sent to rest on another celestial body''~\citep{UA1999}.

When word of the planned tribute reached the Diné, their leaders reacted publicly with dismay and outrage. Then-President of the Navajo Nation Albert Hale was ``appalled and upset'' by the plan. ``It is one thing to prove, to study, to examine and even for men to walk upon the moon. But it is sacrilege, a gross insensitivity to the beliefs of many Native Americans, to place human remains on the moon''~\citep{Volante1998}. Known to the Diné as \emph{Ooljéé’} or \emph{Tł’éé Honaa’éí}, the Moon is ``not merely a celestial body but a deity deeply rooted in Navajo creation stories and ceremonies''~\citep{NavajoNation2024}. Aversion to the remains of the dead is culturally significant for the Diné; in 1997, ``medicine men warned tribe members to stay away from the San Francisco Peaks north of Flagstaff [Arizona, U.S.] after they learned the sacred mountains had been defiled by people scattering cremated remains''~\citep{Volante1998}.

NASA faced unwanted media attention for flying Shoemaker’s ashes to the Moon, much of which focused on the lack of consultation with Native Americans before proceeding with the plan. While it did not rule out the possibility of sending human remains to the Moon in the future, it promised a different course of action would be followed. ``None of the scientists on the program were aware that this would be insensitive,'' said Peggy Wilhide, NASA’s director of public affairs, committing that ``if we ever discuss doing something like this again, we will consult more widely and we will consult with Native Americans''~\citep{Volante1998}. Wilhide’s evident surprise at the notion that Native people might find the act insensitive was echoed by non-Native scientists. Shoemaker’s widow, planetary scientist Carolyn Shoemaker, was ``completely astonished'' at the criticism of NASA. She said that for Gene, the Moon was ``just an important part in his life, and he would never have thought about desecrating it''~\citep{Volante1998}.

Between 1998 and 2021, successive U.S. Administrations set forth new executive policies intended to improve relations between Native American nations and the federal government by engaging in coordination and consultation on matters of cultural importance. In a memorandum of understanding issued during the administration of President Joe Biden, eight major federal agencies agreed ``to work together and consult with Indian Tribes and Native Hawaiian organizations and collaborate with Tribal and Native Hawaiian organization leaders and spiritual leaders, as appropriate, in developing and implementing actions to improve the protection of and access to Tribal and Native Hawaiian sacred sites''~\citep{USDOI2021}. Yet it also noted that the memorandum was ``a voluntary agreement that expresses the good-faith intentions of the Participating Agencies, is not intended to be legally binding, does not create any contractual or fiscal obligations, does not unlawfully extend Federal authority, and is not enforceable by any party.''

As now, there were then no legal constraints on launching human remains to the Moon from U.S. territory nor did there exist any requirement that the federal government consult with Native Americans before conducting such activities. By 2023 commercial activities in space dominated over those conducted exclusively by governments. Astrobotic Technology, an American space company, proposed \emph{Peregrine Mission One} as part of NASA’s Commercial Lunar Payload Services (CLPS) initiative, which allows the agency to outsource the launch and transport of lunar cargo to private companies. NASA paid Astrobotic Technology to carry five payloads on the mission, which was otherwise privately funded. If successful, it would have been the first U.S.-built spacecraft to soft-land on the Moon since the crewed Apollo Lunar Module on \emph{Apollo 17} in 1972.

Two other U.S. companies, Celestis and Elysium Space, were ride-sharers on \emph{Peregrine Mission One}. Both provided what they referred to as ``lunar burial services'', promising clients to deposit small capsules of their loved ones’ cremated remains on the Moon with prices starting at USD 13000~\citep{Bartels2024}. Celestis alone planned to send the remains of 66 people aboard its payload. This upset the Diné, particularly because they were not consulted, contrary to NASA promises made in 1998 and later policy changes. 

In December 2023, Navajo Nation President Buu Nygren addressed his people’s concerns to NASA and the U.S. Department of Transportation (as the regulatory agency responsible for launch clearances via the Federal Aviation Administration). In the letter, Nygren asked the government to delay the launch pending immediate consultations. Nygren wrote, ``We believe that both NASA and the USDOT should have engaged in consultation with us before agreeing to contract with a company that transports human remains to the Moon or authorizing a launch carrying such payloads''~\citep{NavajoNation2023}. Nygren was contacted by NASA Administrator Bill Nelson, who indicated that the agency’s commitment to tribal consultation was unchanged. 

The Biden Administration convened a last-minute meeting with Diné leaders and others to discuss the issue. NASA reportedly told tribal representatives at the meeting it didn’t have to consult with them because the flight wasn’t a NASA mission~\citep{Woodard2024}. Chris Culbert, CLPS Program Manager at NASA's Johnson Space Center, asserted that ``the approval process doesn't run through NASA for commercial missions,'' and therefore the launching companies alone decided what they would launch on the basis of business considerations~\citep{Tingley2024}. Similarly, the Federal Aviation Administration denied any responsibility for the contents of payloads launched into space from U.S. territory, noting that its oversight role is ``statutorily limited to ensuring space flights do not pose a safety or national security threat to the United States.''~\citep{Fisher2025}.

The Diné rejected NASA’s position on the upcoming mission. Nygren said in a contemporaneous interview: ``I told NASA, ‘You promised to consult with us.’ Private companies can get around the rules''~\citep{Rickert2024}. Diné registered objections to the launch through multiple public channels. Justin Ahasteen, Executive Director of the Navajo Nation’s Washington, D.C., office, told CNN that in raising objections to the planned launch the Diné did not assert sovereignty over the Moon, but rather that it only asked for respect: ``We’re turning the moon into a graveyard and we’re turning it into a waste site. At what point are we going to stop and say we need to start protecting the moon as we do the Grand Canyon?''~\citep{Fisher2025}. A public statement released by the Navajo Nation on 4 January included a quotation from Nygren that summarized the view of his people: ``The suggestion of transforming [the Moon] into a resting place for human remains is deeply disturbing and unacceptable to our people and many other tribal nations''~\citep{NavajoNation2024}.

Celestis and Elysium Space publicly criticized Diné leaders. ``We are aware of the concerns expressed by Mr. Nygren, but do not find them substantive,'' the CEO of Celestis, Charles Chafer, said in a statement. Chafer then went on to assail Diné religious freedom. ``No one, and no religion, owns the Moon. If the beliefs of the world’s multitude of religions were considered, it’s quite likely that no missions would ever be approved. Simply put, we do not and never have let religious beliefs dictate humanity’s space efforts. There is not and should not be a religious test''~\citep{Fisher2025}. Astrorobotic Technology CEO John Thornton complained that the Diné did not raise concerns about its launch earlier, pointing out that it announced the mission as early as 2015. ``I’ve been disappointed that this conversation came up so late in the game,'' Thornton said. Yet Thorton maintained ``we really are trying to do the right thing and I hope we can find a good path forward with the Navajo Nation''~\citep{Fisher2025}.

Over continued tribal objections, \emph{Peregrine Mission One} launched on 8 January 2024 during the maiden flight of the Vulcan Centaur rocket. Hours after liftoff, the spacecraft developed a serious fuel leak. It spent six days orbiting the Earth while engineers attempted to correct the problem.  Realizing it could not land the spacecraft as planned, and in order to ensure it would not become a source of space debris, Astrobotic Technology intentionally de-orbited the spacecraft on 18 January. \emph{Peregrine Mission One} and its cargo of human cremated remains re-entered the Earth’s atmosphere over the South Pacific Ocean. The Diné leadership described being ``relieved \ldots that the lunar surface will not become a resting place for human remains'', promising that the Nation ``will continue to object to any spaceflights that include human remains destined for a lunar burial.'' Neither Celestis nor Elysium Space reacted publicly, noting only that elements of the mission were considered by them to be a success, and that they were committed to attempting future missions, which they described as ``a very rare, special, and risky effort''~\citep{Lagatta2024}.

\section{Analysis and best practices for future engagement}
Both incidents took place against the backdrop of repeated broken agreements and treaties between the United States and Native American nations. Through this historical lens, the deposition of human remains on the Moon appears to be another event in the ongoing colonization of the Americas. The lunar cremated remains episodes show that Native Americans offended by the actions of colonizers find themselves without the legal agency to effectively assert their rights insofar as space activities sanctioned by the United States are concerned.

The OST and its conventions anticipate nation-states as the main actors in outer space and not the private actors rapidly dominating space activities. Treaty signatories retain responsibility for supervising the actions of launching entities in order to ensure compliance with Treaty. But the fatal flaw lies in the Treaty itself, framed in such a way as to center a Western capitalist view that considers space an object of human utility subject to exploitation for its benefit. This creates a stark contrast with the frequent Indigenous view of a reciprocal relationship between humans and nature. 

Furthermore, the OST and its implementation process effectively excludes from meaningful participation Indigenous groups that do not possess full legal sovereignty. In the example of \emph{Peregrine Mission One}, the Navajo Nation attempted to work with the U.S. government, respecting that process, yet the government acted as though it had no means of preventing the launch. The difference between the 1998 and 2024 affairs is that in 1998 the action was officially government-sanctioned, whereas in 2024 the government asserted it was a private matter.

The Navajo Nation's call for consultations likely did not invoke the relevant provision in OST Article IX calling for ``due regard'', as the matter was considered internal to the United States. However, it is arguable that the Biden Administration’s actions were at least inconsistent with its stated position with respect to engaging Native American nations. The Navajo Nation, in fact, called the Administration out for failing to adhere to a 2021 memorandum in which the President wrote ``My Administration is committed to honoring Tribal sovereignty and including Tribal voices in policy deliberation that affects Tribal communities''~\citep{Biden2021}.

Professor Michele Hanlon pointed out that ``the Navajo Nation isn’t eligible to join or leave [the OST] on its own but is subject to [it] through the U.S.'' because of its quasi-sovereign status as part of the United States~\citep{Bartels2024}. This makes the federal government the representative of the Navajo Nation to the United Nations Committee on the Peaceful Uses of Outer Space (COPUOS), but it is not bound to articulate the Diné point of view on any particular matter of space law. With a tribal membership representing only 0.12\% of the population of the U.S., it is difficult to envision any scenario in which the Diné wield sufficient political power to influence the U.S. position on space policy matters through the democratic process. 

This, of course, does not change the fact that even if Diné claims under the OST were subject to adjudication through COPUOS or in other venues, there is no clear penalty or sanction in case OST provisions are violated. Signatories can simply ignore violations and excuse their own behavior, as implementation of the OST is based solely on norms of behavior and a consensus process with minimal United Nations oversight. Were the Navajo Nation a truly sovereign entity, it might bring a claim against the United States at the International Court of Justice, but the U.S. withdrew from compulsory jurisdiction in 1986 to accept the Court's jurisdiction only on a discretionary basis. While an independent Navajo Nation could become a State Party to the OST, even in that hypothetical situation the U.S. would only owe it consultation under Article IX and not any particular result from the consultation. The status of Indigenous claims is likewise murky under the U.S.-led Artemis Accords, which some countries view as a positive step toward codifying key principles of space law while others suspect are an outlet for the United States to impose its own quasi-legal rules through a distinctly American interpretation of the OST~\citep{Newman2020}.

Despite all this, there are arguments in favor of intrinsic rights for Indigenous peoples where outer space is concerned. Although it does not mention outer space explicitly, Neilson~\citeyearpar[2--4]{Neilson2024b} points out that many provisions of the United Nations Declaration of the Rights of Indigenous Peoples indirectly address both cultural and economic rights with respect to space. He further suggests that an Indigenous approach to the problem of future uses of outer space ``as seeking to not treat outer space as a resource for the benefit or profit of humanity, but, instead, that we live in treaty with outer space as its own land and place with its own rights of existence''~\citep[6]{Neilson2024a}.

Future approaches to space policy, and in particular policies focusing on human activities on the Moon, could be decolonized in part by bringing sovereign Indigenous peoples fully into the policy-making process. But to the extent that the main object of international space policy, the OST, is a fundamentally colonial instrument, it is difficult to envision effective results as long as it persists in its current form. A new view of the human role in outer space is required, one in which treaties are not made solely between representatives of human groups but rather between humans and space itself. Admittedly such thinking is broadly antithetical to the Western dominionist view of nature and its relationship to humans. Yet recent decades have seen important strides toward decolonizing terrestrial legal systems that offer useful lessons to those seeking to extend these gains to space~\citep{BinagwahoFreeman2021,Shawush2022}.

With this in mind, we conclude with thoughts on appropriate ways to engage Indigenous nations in co-creating a human future in space:
\begin{itemize}
\item First, much of that future is preconditioned on a wider recognition of the inherent rights of Indigenous peoples, a struggle that continues around the world. Until Indigenous nations achieve substantial sovereignty over their own affairs, their participation in space policy discussions cannot be full.
\item Second, meaningful engagement in decision making must accept Indigenous methodologies as distinct and equally valid alongside other ways of knowing. Rejection of inherently decolonizing views denies proper representation to all people. 
\item Third, consultation alone does not constitute meaningful engagement with process. Given the consensus nature of international norms and standards for acceptable actions undertaken in space, including on the Moon, anything short of free, prior and informed consent is inconsistent with the idea enshrined in the Outer Space Treaty that space is ``the province of all mankind''.
\end{itemize}

\bibliographystyle{iaulike} 
{\raggedright  
\bibliography{Barentine-IAUS-399}
}              

\end{document}